\documentclass[12pt,twoside]{article}

\evensidemargin=\oddsidemargin
\def\Title#1#2#3{%
    \baselineskip=18pt
    \begin{center}
          {\large\bf{#1} \\ }
          \bigskip\bigskip
          {#2} \\
          {#3} \\
    \end{center}}
\long\def\Abstract#1{%
         \bigskip
         \parbox{0.93\textwidth}{%
                 \begin{center}
                       {\bf Abstract} \\
                 \end{center}
                 \medskip{\baselineskip=14pt #1}
                 \vss}
         \bigskip}

\makeatletter
\renewcommand{\section}%
 {\@startsection{section}{1}{0pt}%
  {-3.25ex plus -1ex minus -.2ex}{1.5ex plus .2ex}%
  {\vspace*{5mm}\raggedright\large\bf }}
\renewcommand{\thesection}{\arabic{section}.}
\@addtoreset{equation}{section}
\renewcommand{\@eqnnum}{(\thesection\theequation)}
\renewcommand{\p@equation}{\thesection}
\makeatother

\begin{document}

\Title{Hamiltonian formulation of General Relativity\\
50 years after the Dirac celebrated paper:\\
do unsolved problems still exist?}%
{T. P. Shestakova}%
{Department of Theoretical and Computational Physics,
Southern Federal University\footnote{former Rostov State University},\\
Sorge St. 5, Rostov-on-Don 344090, Russia \\
E-mail: {\tt shestakova@sfedu.ru}}

\Abstract{About 50 years ago, in 1958, Dirac published his formulation of generalized Hamiltonian dynamics for gravitation. Several years later Arnowitt, Deser and Misner (ADM) proposed their description of the dynamics of General Relativity which became a basis of the Wheeler - DeWitt Quantum Geometrodynamics. There exist also other works where the Hamiltonian formulation of gravitational theory was discussed. In spite of decades passed from the famous papers by Dirac and ADM, there are unsolved problems. Namely, are the Dirac and ADM formulations equivalent to each other? Are these formulations equivalent to the original (Lagrangian) Einstein theory? Is the group of transformation in phase space the same as the group of gauge transformation of the Einstein theory? What are rules according to which a generator of transformations in phase space should be constructed? Let us mention also another approach based on extended phase space where gauge degrees of freedom are treated on the equal ground with physical degrees of freedom. Our purpose is to review the above questions and to demonstrate advantages of the extended phase space approach by the example of a simple model with finite number degrees of freedom.}

\section{Introduction}
In 2008 fifty years passed after the publication of the Dirac famous paper, devoted to Hamiltonian form of the theory of gravitation \cite{Dirac1}. Dirac was not tired of repeating that "any dynamical theory must first be put in the Hamiltonian form before one can quantize it". However, when constructing the Hamiltonian, Dirac made an additional assumption that $g_{0i}=0$ which noticeably simplified his calculations but also led him to the conclusion that "this simplification can be achieved only at the expense of abandoning four-dimensional symmetry". It may have become a reason why the formulation by Dirac was not so recognized as the one by Arnowitt, Deser and Misner (ADM) where the parametrization of gravitational variables through lapse and shift functions was introduced \cite{ADM}. It is needless to recall that the ADM formulation with its clear geometrical interpretation became a basis of the Wheeler - DeWitt Quantum Geometrodynamics which serves as a theoretical foundation for many works exploring the Early Universe.

Meanwhile, in 2008 the paper by Kiriushcheva and Kuzmin appeared \cite{KK}, where the authors claim that the formulation by Dirac and that by ADM are not equivalent since these two formulations are related by a non-canonical transformation of phase space variables from the 4-metrics $g_{\mu\nu}$ to the lapse and shift functions $N$, $N_i$ and the 3-metrics $\gamma_{ij}$. The authors demonstrate by direct calculations that the Dirac assumption $g_{0i}=0$ does not affect the form of the total Hamiltonian and thus is not necessary at all for constructing the Hamiltonian formulation. If so, we face the question what formalism should be chosen. Keeping in mind that the Dirac formulation deals with original gravitational variables, it seems to be more preferable. The second argument in favor of it is that it seems to be possible to construct a generator of transformations in phase space which gives correct diffeomorphism transformations for the 4-metrics $g_{\mu\nu}$, while a well-known shortcoming of the ADM formulation is that the derivation of correct transformations for the lapse and shift functions (corresponding to diffeomorphism transformations of the metrics) is rather problematic.

Therefore, the authors of the paper \cite{KK} have raised a serious problem: Should we abandon the ADM formulation because the new ADM variables are not related with the old ones by a canonical transformation? Should we also abandon any other parametrizations if they do not satisfy this criterion? Let us mention that from the very beginning the complexity of General Relativity inspired the search for suitable parametrizations to solve various problems. There are other examples of constructing Hamiltonian dynamics of General Relativity, for instance, the formulation by Faddeev in his work devoted to gravitational energy \cite{Faddeev}. He made use of the following variables
\begin{equation}
\label{Fadd}
h^{\mu\nu}=\sqrt{-g}g^{\mu\nu};\quad
\lambda^0=\frac1{h^{00}}+1;\quad
\lambda^i=\frac{h^{0i}}{h^{00}};\quad
q^{ij}=h^{0i}h^{0j}-h^{00}h^{ij}.
\end{equation}
As far as I know, the canonicity of transformations to new variables was not proved in the most of cases. Then, should we consider the ADM formulation as a model named, say, "ADM gravity", without any reference to Einstein General Relativity, as Kiriushcheva and Kuzmin suggest?  They declared that in their analysis they follow to what Lagrange called "regular rules of procedure". We shall start our consideration from the question, what rules of procedures one ought to take into account when constructing a Hamiltonian formulation of gravitational theory.

\section{``The rules of procedure'': generalized Hamiltonian dynamics}
Firstly, let us turn to generalized Hamiltonian dynamics of a theory with constraints. It is based upon several rules postulated by Dirac that can be verified only by results of their applications to real physical fields. The postulates, which are of interest for us, are the following:

1. The Hamiltonian of the theory should be constructed according to the rule
\begin{equation}
\label{gen.Ham}
H=p_a\dot q^a-L+\lambda_a\varphi^a,
\end{equation}
where $p_a$, $q^a$ are pairs of variables called canonical in the sense that all the velocities $\dot q^a$ can be expressed through conjugate momenta; $\varphi^a$ is the set of constraints and $\lambda_a$ are Lagrange multipliers. In the case of gravitational theory the Hamiltonian is
\begin{equation}
\label{grav.Ham}
H_G=p^{ij}\dot g_{ij}-L.
\end{equation}
There are modifications of this rule, some authors include into the form $p_a\dot q^a$ also non-canonical variables in the above sense, i.e. those for which it would be impossible to express the velocities in terms of the momenta. Then we have the total Hamiltonian which for gravity takes the form
\begin{equation}
\label{tot.Ham}
H_T=p^{00}\dot g_{00}+p^{0i}\dot g_{0i}+H_G.
\end{equation}
Since $p^{00}=0$ are primary constraints, the two Hamiltonians coincide on the constraints surface, and the difference between them seems not to be of importance. However, as we shall see, the difference becomes essential when constructing a generator of gauge transformations. Let us note that Dirac in his work \cite{Dirac1} obtained the canonical Hamiltonian (\ref{grav.Ham}) while Kiriushcheva and Kuzmin \cite{KK} considered the total Hamiltonian (\ref{tot.Ham}). In any case, making use of the total Hamiltonian implies {\it a mixed formalism} in which the Hamiltonian is written in terms of canonical coordinates and momenta but as well of velocities that cannot be expressed through the momenta.

2. The constraints or their linear combinations play the role of generators of gauge transformations \cite{Dirac2}. Dirac considered the theory of electromagnetic field as a typical example of a field theory with constraints. Indeed, in this theory the secondary constraint produces correct transformations of spatial components of vector potential of electromagnetic field $A_i$. However, even in this simple case the constraints do not generate the transformation of the zero component of vector potential $A_0$. The same situation we face in gravitational theory: while the transformations of the 3-metrics can be generated by the secondary (momentum) constraints, the transformation for $g_{0\mu}$ cannot be obtained in this way. To avoid this difficulty, some other algorithms were suggested \cite{Cast,BRR} which were thoroughly discussed in \cite{KK}.

The methods \cite{Cast,BRR} give the correct transformations for the original gravitational variables $g_{\mu\nu}$, used by Dirac, and do not lead to the correct transformations for the ADM variables. As has been already mentioned, it can be considered as an additional argument why one should abandon the ADM dynamics. However, would not it better to think, what are the reasons that prevent us from obtaining correct transformations in the ADM case? In Section 3 we shall see that the rules of constructing the generator are rather artificial and hardly can be recognized as fundamental ``rules of procedure'' in the sense of Lagrange. We do not even have a strong criterion what exact form a Hamiltonian must take for a constrained theory. It returns us to the questions: Are the Dirac and ADM formulations really non-equivalent to each other? What are rules according to which a generator of transformations in phase space should be constructed? Is there an alternative to the both approaches? In Section 4 I shall touch the extended phase space approach that may present such an alternative, and I shall try to answer these questions in Conclusion.

\section{``The rules of procedure'': the generator of gauge transformations}
In \cite{Cast} the generator of gauge transformations is sought in the form
\begin{equation}
\label{gen.gen}
G=\sum\limits_n\theta_{\mu}^{(n)}G_n^{\mu},
\end{equation}
where $G_n^{\mu}$ are first class constraints, $\theta_{\mu}^{(n)}$ are the $n$th order time derivatives of the gauge parameters $\theta_{\mu}$. In other words, from the very beginning it is taken into account that variations of canonical variables may depend on time derivatives of $\theta_{\mu}$. In the theory of gravity the variations of $g_{\mu\nu}$ involve first order derivatives of gauge parameters, thus the generator is
\begin{equation}
\label{grav.gen}
G=\theta_{\mu}G_0^{\mu}+\dot\theta_{\mu}G_1^{\mu}.
\end{equation}
$G_n^{\mu}$ satisfy the following conditions that were derived from the requirement of invariance of motion equations under transformations in phase space:
\begin{equation}
\label{gen.cond1}
G_1^{\mu}\quad{\rm are\;primary\;constraints};
\end{equation}
\begin{equation}
\label{gen.cond2}
G_0^{\mu}+\left\{G_1^{\mu},\;H\right\}\quad{\rm are\;primary\;constraints};
\end{equation}
\begin{equation}
\label{gen.cond3}
\left\{G_0^{\mu},\;H\right\}\quad{\rm are\;primary\;constraints}.
\end{equation}
In \cite{KK} the generator (\ref{grav.gen}) was calculated for the full gravitational theory. Paying tribute to the authors of the paper \cite{KK} for their huge work of making cumbersome calculations, I shall demonstrate, nevertheless, that basic rules can be understood taking a simple model as an example. Consider an isotropic cosmological model with the Lagrangian
\begin{equation}
\label{Lagr1}
L=-\frac12\frac{a\dot a^2}N+\frac12 Na.
\end{equation}
This model is traditionally described in the ADM variables ($N$ is the lapse function, $a$ is the scale factor). For our purpose, it is more convenient to go to a new variable $\mu=N^2$ which corresponds to $g_{00}$. So the Lagrangian is
\begin{equation}
\label{Lagr2}
L=-\frac12\frac{a\dot a^2}{\sqrt{\mu}}+\frac12\sqrt{\mu}\,a.
\end{equation}
The canonical Hamiltonian corresponding to (\ref{grav.Ham}) reads
\begin{equation}
\label{Ham1}
H_G=-\frac12\frac{\sqrt{\mu}}a\;p^2-\frac12\sqrt{\mu}\,a
\end{equation}
and the total Hamiltonian (see (\ref{tot.Ham})) is
\begin{equation}
\label{Ham2}
H_T=\pi\dot\mu-\frac12\frac{\sqrt{\mu}}a\;p^2-\frac12\sqrt{\mu}\,a
\end{equation}
($p$ is the momentum conjugate to the scale factor, $\pi$ is the momentum conjugate to the gauge variable $\mu$). $\pi=0$ is the only primary constraint of the model, so that
\begin{equation}
\label{cond1}
G_1=\pi.
\end{equation}
The secondary constraint is
\begin{equation}
\label{sec.constr1}
\dot\pi=\left\{\pi,\;H_T\right\}=-\frac{\partial H_T}{\partial\mu}
 =\frac14\frac1{a\sqrt{\mu}}\;p^2+\frac14\frac a{\sqrt{\mu}}=T.
\end{equation}
The canonical Hamiltonian (\ref{Ham1}) appears to be proportional to the secondary constraint $T$, $H_G=-2\mu T$.

The condition (\ref{gen.cond2}) becomes
\begin{equation}
\label{cond2}
G_0+\left\{\pi,\;H_T\right\}=\alpha\pi;
\end{equation}
\begin{equation}
\label{cond2_1}
G_0=-T+\alpha\pi,
\end{equation}
$\alpha$ is a coefficient that can be found from the requirement (\ref{gen.cond3}):
\begin{equation}
\label{cond3}
\left\{G_0,\;H_T\right\}=\beta\pi;
\end{equation}
\begin{eqnarray}
\left\{G_0,\;H_T\right\}
&=&-\left\{T,\;H_T\right\}+\alpha\left\{\pi,\;H_T\right\}=\nonumber\\
\label{cond3_1}
&=&-\left\{T,\;\pi\dot\mu-2\mu T\right\}+\alpha T
  =-\left\{T,\;\pi\right\}\dot\mu+\alpha T
  =\frac1{2\mu}\;\dot\mu T+\alpha T;
\end{eqnarray}
\begin{equation}
\label{betaal}
\beta=0;\quad
\alpha=-\frac1{2\mu}\;\dot\mu;
\end{equation}
\begin{equation}
\label{cond2_2}
G_0=-\frac1{2\mu}\;\dot\mu\pi-T.
\end{equation}
The full generator $G$ (\ref{grav.gen}) can be written as
\begin{equation}
\label{gen1}
G=\left(-\frac1{2\mu}\;\dot\mu\pi-T\right)\theta+\pi\dot\theta.
\end{equation}
The transformation of the variable $\mu$ is
\begin{equation}
\label{mu_transf}
\delta\mu=\left\{\mu,\;G\right\}
 =-\frac1{2\mu}\;\dot\mu\theta+\dot\theta.
\end{equation}
The same expression (up to the multiplier being equal to 2) can be obtained from general transformations of the metric tensor,
\begin{equation}
\label{g_transf}
\delta g_{\mu\nu}
 =\theta^{\lambda}\partial_{\lambda}g_{\mu\nu}
 +g_{\mu\lambda}\partial_{\nu}\theta^{\lambda}
 +g_{\nu\lambda}\partial_{\mu}\theta^{\lambda};
\end{equation}
\begin{equation}
\label{g00_transf}
\delta g_{00}
 =\dot g_{00}\theta^0+2g_{00}\dot\theta^0,
\end{equation}
if one keeps in mind that $g_{00}=\mu$ and in the above formulas $\theta=\theta_0=g_{00}\theta^0$.

It is not difficult to see why the method does not work for the Lagrangian (\ref{Lagr1}). Indeed, in the last case the total Hamiltonian is
\begin{equation}
\label{Ham3}
H_T=\pi\dot N-\frac12\frac Na\;p^2-\frac12\;N\,a
\end{equation}
Again, $\pi$ is the momentum conjugate to the gauge variable $N$, and $\pi=0$ is the only primary constraint. Now the secondary constraint does not depend on $N$:
\begin{equation}
\label{sec.constr2}
\dot\pi=\left\{\pi,\;H_T\right\}=-\frac{\partial H_T}{\partial N}
 =\frac1{2a}\;p^2+\frac12\; a=T,
\end{equation}
therefore, the Poisson bracket $\left\{T,\;\pi\right\}$ in (\ref{cond3_1}) is equal to zero, and one would obtain an incorrect expression for the generator,
\begin{equation}
\label{gen2}
G=-T\theta+\pi\dot\theta,
\end{equation}
which cannot produce the correct variation of $N$,
\begin{equation}
\label{N_transf}
\delta N=-\dot N\theta-N\dot\theta,
\end{equation}
Obviously, the algebra of constraints is not invariant under the choice of parametrization. One can come to the same conclusion by applying the other approach \cite{BRR,KK} of constructing the generator. It is suggested in this approach that the generator ought to be of the form
\begin{equation}
\label{gen3}
G=\eta_{\mu}\pi^{\mu}+\theta_{\mu}T^{\mu},
\end{equation}
$\eta_{\mu}$ and $\theta_{\mu}$ are the two sets of parameters that correspond to primary $\pi^{\mu}$ and secondary $T^{\mu}$ constraints. The relations between $\eta_{\mu}$ and $\theta_{\mu}$ are derived from the requirement of commutativity of variation and differentiation with respect to time of generalized coordinates. These relations can be obtained from Hamiltonian equations of motion and rely upon the algebra of constraints. Thus, it is not surprised at all, that this approach leads to the same results as the method \cite{Cast}. One can say that the difference in the algebra of constraints is a consequence of the fact that the two parametrizations used above are not related by a canonical transformation. However, such a transformation has to involve gauge variables like $\mu$, $N$, which in the original Dirac approach play the role of Lagrangian multipliers at constraints and are not included into the set of canonical variables. To treat these variables on the equal basis with the others, {\it one should extend the phase space}.

We would also like to emphasize that the correct expression (\ref{gen1}) is a consequence of making use of the total Hamiltonian (\ref{tot.Ham}) instead of the canonical one (\ref{grav.Ham}). As one can see from (\ref{cond3_1}), when applying the canonical Hamiltonian one would obtain the incorrect expression (\ref{gen2}). However, {\it the total Hamiltonian (\ref{tot.Ham}) has been already defined in extended phase space}. Therefore, {\it the derivation of the correct generator requires abandoning the original rules prescribed by Dirac and replacing them by a new procedure} presented in \cite{Cast}. Nevertheless, this new procedure gives right results for a certain parametrization of gravitational variables only.

\section{Extended phase space}
The idea of extended phase space appeared in the works by Batalin, Fradkin and Vilkovisky (BFV) where their approach to path integral quantization of gauge theories was proposed \cite{BFV1,BFV2,BFV3}. The Hamiltonian form of the BFV effective action contains the part $p_a\dot q^a$, that comprises all pair of conjugate variables including gauge ones, the Dirac canonical Hamiltonian with linear combinations of constrains, as well as gauge-fixing and ghost parts. The transformations of phase variables are generated by the BRST charge which is constructed as a series in powers of Grassmannian variables with coefficients given by generalized structure functions:
\begin{equation}
\label{gen_BFV}
\Omega=c^{\alpha}U^{(0)}_{\alpha}
 +c^{\beta}c^{\gamma}U^{(1)\alpha}_{\gamma\beta}\bar\rho_{\alpha}+\ldots
\end{equation}
The structure functions of zero order are the full set of primary and secondary constraints, the structure functions of high orders are determined by the algebra of constraints. For the model with the Lagrangian (\ref{Lagr1}) the algebra of constraints is obviously Abelian, and the consequent application of the BFV approach would lead us to the same incorrect expression (\ref{gen2}).

For the full gravitational theory the structure functions of the first order are not equal to zero \cite{Shest}, so that the BRST generator involves a term with three ghost fields. This generator also cannot produce correct transformations for all gravitational variables as one can check considering the model with the Lagrangian (\ref{Lagr2}). In this case the only non-zero structure function of the first order is
\begin{equation}
\label{str_fun}
\left\{\pi,\; T\right\}=\frac1{2\mu}\; T=C^2_{\; 12}T;\quad
U^{(1)2}_{12}=-\frac12\; C^2_{\; 12}=-\frac1{4\mu}.
\end{equation}

The BFV approach aimed to reproduce the Dirac procedure of quantization of theories with constraints at path integral level, and it indeed leads to equivalent results for Yang - Mills theories and some simple models. However, it meets certain difficulties for the full gravitational theory or arbitrary parametrization of gravitational variables. This makes us search for another way of constructing Hamiltonian dynamics in extended phase space.

Such a way has been proposed in our papers \cite{SSV1,SSV2}. A serious shortcoming of the total Hamiltonian (\ref{tot.Ham}) is that it contains generalized velocities apart from coordinates and momenta. To compensate them, one can introduce the missing velocities into the Lagrangian by means of gauge conditions in differential form. For the model (\ref{Lagr1}) we have
\begin{equation}
\label{gfix}
N=f(a)\quad\Rightarrow\quad
\dot N=\frac{df}{da}\;\dot a.
\end{equation}
To give a full consideration, one should also include the ghost sector into the model,
\begin{equation}
\label{Lghost}
L_{(ghost)}=\dot{\bar\theta}N\dot\theta
 +\dot{\bar\theta}\left(\dot N-\frac{df}{da}\;\dot a\right)\theta,
\end{equation}
so that
\begin{eqnarray}
L&=&-\frac12\frac{a\dot a^2}N+\frac12 Na
 +\lambda\left(\dot N-\frac{df}{da}\;\dot a\right)
 +\dot{\bar\theta}\left(\dot N-\frac{df}{da}\;\dot a\right)\theta
 +\dot{\bar\theta}N\dot\theta=\nonumber\\
\label{Lagr3}
&=&-\frac12\frac{a\dot a^2}N+\frac12 Na
 +\pi\left(\dot N-\frac{df}{da}\;\dot a\right)
 +\dot{\bar\theta}N\dot\theta.
\end{eqnarray}
The conjugate momenta are:
\begin{equation}
\label{mom1}
\pi=\lambda+\dot{\bar\theta}\theta;\quad
p=-\frac{a\dot a}N-\pi\frac{df}{da};\quad
\bar{\cal P}=N\dot{\bar\theta};\quad
{\cal P}=N\dot\theta.
\end{equation}

Let us now go to a new variable
\begin{equation}
\label{ch_var1}
N=v\left(\tilde N,\; a\right).
\end{equation}
At the same time, the rest variables are unchanged:
\begin{equation}
\label{ch_var2}
a=\tilde a;\quad
\theta=\tilde\theta;\quad
\bar\theta=\tilde{\bar\theta}.
\end{equation}
It is the analog of the transformation from the original gravitational variables $g_{\mu\nu}$ to the ADM variables. Indeed, in the both cases only gauge variables are transformed while the rest variables remain unchanged. It was shown in \cite{KK} that such a transformation is not canonical. The reason is that the momenta conjugate to physical variables also remain unchanged. The situation in extended phase space is different. After the change (\ref{ch_var1}) the Lagrangian is written as
\begin{equation}
\label{Lagr4}
L=-\frac12\frac{a\dot a^2}{v\left(\tilde N,\; a\right)}
 +\frac12\;v\left(\tilde N,\; a\right)a
 +\pi\left(\frac{\partial v}{\partial\tilde N}\;\dot{\tilde N}
  +\frac{\partial v}{\partial a}\;\dot a
  -\frac{df}{da}\;\dot a\right)
 +v\left(\tilde N,\; a\right)\dot{\bar\theta}\dot\theta.
\end{equation}
The new momenta are:
\begin{equation}
\label{mom2_1}
\tilde\pi=\pi\frac{\partial v}{\partial\tilde N};
\end{equation}
\begin{equation}
\label{mom2_2}
\tilde p=-\frac{a\dot a}{v\left(\tilde N,\; a\right)}
  +\pi\frac{\partial v}{\partial a}-\pi\frac{df}{da}
 =p+\pi\frac{\partial v}{\partial a};
\end{equation}
\begin{equation}
\label{mom2_3}
\tilde{\bar{\cal P}}=v\left(\tilde N,\; a\right)\dot{\bar\theta}
 =\bar{\cal P};
\end{equation}
\begin{equation}
\label{mom2_4}
\tilde{\cal P}=v\left(\tilde N,\; a\right)\dot\theta={\cal P}.
\end{equation}
It is easy to demonstrate that the transformations (\ref{ch_var1}), (\ref{ch_var2}), (\ref{mom2_1}) -- (\ref{mom2_4}) are canonical in extended phase space. The generating function will depend on new coordinates and old momenta \cite{LL},
\begin{equation}
\label{gen.fun}
\Phi\left(\tilde N,\;\tilde a,\;\tilde{\bar\theta},\;\tilde\theta,\;
  \pi,\;p,\;\bar{\cal P},\;{\cal P}\right)
 =-\pi\, v\left(\tilde N,\;\tilde a\right)
  -p\,\tilde a-\bar{\cal P}\,\tilde\theta
  -\tilde{\bar\theta}\,{\cal P}.
\end{equation}
Then the relations
\begin{equation}
\label{tr1}
N=-\frac{\partial\Phi}{\partial\pi};\quad
a=-\frac{\partial\Phi}{\partial p};\quad
\tilde\pi=-\frac{\partial\Phi}{\partial\tilde N\vphantom{\sqrt N}};\quad
\tilde p=-\frac{\partial\Phi}{\partial\tilde a};
\end{equation}
\begin{equation}
\label{tr2}
\theta=-\frac{\partial\Phi}{\partial\bar{\cal P}\vphantom{\sqrt N}};\quad
\bar\theta=-\frac{\partial\Phi}{\partial{\cal P}};\quad
\tilde{\cal P}=-\frac{\partial\Phi}{\partial\tilde{\bar\theta}};\quad
\tilde{\bar{\cal P}}=-\frac{\partial\Phi}{\partial\tilde\theta}
\end{equation}
give exactly the transformation (\ref{ch_var1}), (\ref{ch_var2}), (\ref{mom2_1}) -- (\ref{mom2_4}). On the other hand, one can check that Poisson brackets among all phase variables maintain their canonical form.

It is well known that, while introducing a gauge condition breaks down local gauge invariance, there remains global BRST invariance. And it is remarkable that the existing of BRST invariance enables us to construct the generator of transformations in extended phase space (BRST charge) making use of the first Noether theorem. As a consequence of a global symmetry, there exists a conserved quantity which can be obtained from equations of motion. For Lagrangians without derivatives of high orders, the BRST generator is
\begin{equation}
\label{gen.BRST}
\Omega=\int\!d^3x\,\frac{\partial L}{\partial q^a}\,\delta q^a
 =\int\!d^3x\,p_a\delta q^a,
\end{equation}
where $p_a$ are generalized momenta and $\delta q^a$ are variations of generalized coordinates under which the action remains unchanged. Obviously, the BRST charge generates correct transformations if the theory is not degenerate, i.e. derivatives of the Lagrangian with respect to velocities are not zero:
\begin{equation}
\label{BRST_tr}
\delta q^a=\left\{q^a,\;\Omega\right\}
 =\frac{\delta\Omega}{\delta p_a}.
\end{equation}
The extension of phase space removes degeneracy of the theory. Another necessary condition of consistency of this approach is a complete equivalence between the Lagrangian formulation and the Hamiltonian dynamics in extended phase space that allows one to write down the BRST charge in terms of coordinates and momenta. This equivalence is ensured by construction of the Hamiltonian dynamics itself and was demonstrated in our works \cite{SSV1,SSV2}.

For the model considered in the present paper the Hamiltonian in extended phase space for our model looks like
\begin{eqnarray}
\tilde H
 &=&-\frac12\;\frac{v\left(\tilde N,\; a\right)}a\left[\tilde p^2
    +2\tilde p\tilde\pi\;
     \frac1{\displaystyle\frac{\partial v}{\partial\tilde N\vphantom{\sqrt N}}}\;\frac{df}{da}
    +\tilde\pi^2\;
     \frac1{\left(\displaystyle\frac{\partial v}{\partial\tilde N\vphantom{\sqrt N}}\right)^2}\;
     \left(\frac{df}{da}\right)^2-\right.\nonumber\\
 &-&\left.2\tilde p\tilde\pi\;
     \frac1{\displaystyle\frac{\partial v}{\partial\tilde N\vphantom{\sqrt N}}}\;
     \frac{\partial v}{\partial a}
    -2\tilde\pi^2\;
     \frac1{\left(\displaystyle\frac{\partial v}{\partial\tilde N\vphantom{\sqrt N}}\right)^2}\;
     \frac{\partial v}{\partial a}\;\frac{df}{da}
    +\tilde\pi^2\;
     \frac1{\left(\displaystyle\frac{\partial v}{\partial\tilde N\vphantom{\sqrt N}}\right)^2}\;
     \left(\frac{\partial v}{\partial a}\right)^2\right]-\nonumber\\
\label{Ham.EPS}
  &-&\frac12\; v\left(\tilde N,\; a\right)a
   +\frac1{v\left(\tilde N,\; a\right)}\;\tilde{\bar{\cal P}}\tilde{\cal P}
\end{eqnarray}
and the BRST generator is
\begin{equation}
\label{BRST_mod}
\Omega=-\tilde H\theta
 -\frac1{\displaystyle\frac{\partial v}{\partial\tilde N\vphantom{\sqrt N}}}\;\tilde\pi{\cal P}.
\end{equation}

One can check that $\Omega$ (\ref{BRST_mod}) generates the correct transformations for any gauge variable $\tilde N$ given by the relation (\ref{ch_var1}),
\begin{equation}
\label{tilN_trasf}
\delta\tilde N=\left\{\tilde N,\;\Omega\right\}
 =-\frac{\partial\tilde H}{\partial\tilde\pi}\;\theta
  -\frac1{\displaystyle\frac{\partial v}{\partial\tilde N\vphantom{\sqrt N}}}\;{\cal P}
 =-\dot{\tilde N}\theta
  -\frac1{\displaystyle\frac{\partial v}{\partial\tilde N\vphantom{\sqrt N}}}\;
   v\left(\tilde N,\; a\right)\;\dot\theta.
\end{equation}
(It is taken into account that the gauge condition is included into the set of Hamiltonian equations in extended phase space, $\dot{\tilde N}=\displaystyle\frac{\partial\tilde H}{\partial\tilde\pi}$.) In particular, for the original variable $N$ one gets the correct transformation (\ref{N_transf}).

\section{Conclusions}
From the viewpoint of the Lagrangian formalism, the original parametrization of gravitational variables and the ADM parametrization are completely equivalent. Nothing in the Lagrangian formalism prevents from making use of various parametrizations. Therefore, non-equivalence of any two Hamiltonian formulations of the theory witnesses some problem in the construction of Hamiltonian formalism. In our extended phase space approach we do not need to abandon generally accepted rules of constructing a Hamiltonian form of the theory or invent some new rules. Indeed, in our approach
\begin{itemize}
\item the Hamiltonian is built up according to the usual rule $H=p_a\dot q^a-L$;
\item the Hamiltonian equations in extended phase space are completely equivalent to the Lagrangian equations;
\item due to global BRST invariance it appears to be possible to construct the BRST charge in conformity with the
first Noether theorem which produces correct transformations for all phase variables.
\end{itemize}
The only additional assumption we have made is introducing into the Lagrangian missing velocities by means of the differential form of gauge conditions.

Of course, until now our results have been demonstrated for simple models and it would be important to reproduce them for the full gravitational theory. However, the results are rather indicative, so I cannot see any reason to abandon the ADM parametrization or other possible parametrizations. If one used the formalism of extended phase space, the formulation in the ADM variables is expected to be equivalent to the Hamiltonian formulation in terms of metric tensor and conjugate momenta, as well as to the original Einstein General Relativity. In this case the group of transformation in phase space includes the group of gauge transformation of the Einstein theory, and we have an explicit prescription how to build up the generator of these transformations. The proposed formulation in extended phase space reveals new prospects to quantization of gravity, as was reported at previous PIRT meetings.


\small

\begin{thebibliography}{99}
\itemsep=-5pt
\bibitem{Dirac1}
P. A. M. Dirac,
 {\it Proc. Roy. Soc.\/} {\bf A246} (1958), P. 333--343.
\bibitem{ADM}
R. Arnowitt, S. Deser and C. W. Misner,
 ``The Dynamics of General Relativity'',
 in: {\it Gravitation, an Introduction to Current Research\/},
 ed. by L. Witten, John Wiley \& Sons, New York (1963), P. 227--284.
\bibitem{KK}
N. Kiriushcheva and S. V. Kuzmin,
 ``The Hamiltonian formulation of General Relativity: myth and reality'',
 E-print arXiv: gr-qc/0809.0097.
\bibitem{Faddeev}
L. D. Faddeev,
 ``The energy problem in Einstein's theory of gravitation'',
 {\it Usp. Fiz. Nauk\/} {\bf 136} (1982), P. 435--457.
\bibitem{Dirac2}
P. A. M. Dirac,
 {\it Lectures on Quantum Mechanics\/}, Yeshiva University, New York (1964).
\bibitem{Cast}
L. Castellani,
 {\it Ann. Phys.\/} {\bf 143} (1982), P. 357--371.
\bibitem{BRR}
R. Banerjee, H. J. Rothe and K. D. Rothe,
 {\it Phys. Lett.\/} {\bf B463} (1999), P. 248--251.
\bibitem{BFV1}
E. S. Fradkin and G. A. Vilkovisky,
 {\it Phys. Lett.\/} {\bf B55} (1975), P. 224--226.
\bibitem{BFV2}
I. A. Batalin and G. A. Vilkovisky,
 {\it Phys. Lett.\/} {\bf B69} (1977), P. 309--312.
\bibitem{BFV3}
E. S. Fradkin and T. E. Fradkina,
 {\it Phys. Lett.\/} {\bf B72} (1978), P. 343--348.
\bibitem{Shest}
T. P. Shestakova,
 {\it Path integral approach to Quantum Field Theory\/},
 Moscow - Izhevsk (2005) (in Russian).
\bibitem{SSV1}
V. A. Savchenko, T. P. Shestakova and G. M. Vereshkov,
 {\it Gravitation \& Cosmology\/} {\bf 7} (2001), P. 18--28.
\bibitem{SSV2}
V. A. Savchenko, T. P. Shestakova and G. M. Vereshkov,
 {\it Gravitation \& Cosmology\/} {\bf 7} (2001), P. 102--116.
\bibitem{LL}
L. D. Landau and E. M. Lifshits,
 {\it Mechanics\/}, Moscow (1988).
\end{thebibliography}
\end{document}